\documentclass[amsmath,amssymb,showpacs,twocolumn]{revtex4}
\usepackage{graphicx}
\usepackage{dcolumn}
\usepackage{bm}
\usepackage{color}

\usepackage{amscd,amsmath,amsthm,amsfonts,amssymb}

\def\sech{{\rm sech}}

\def\Imag{{\rm Im}}

\def\ri{{\rm i}}
\def\rd{{\rm d}}
\def\re{{\rm e}}

\begin{document}
\title{Light propagation in periodically modulated complex waveguides}
\author{Sean Nixon}
\author{Jianke Yang} \email{jxyang@uvm.edu}
\affiliation{Department of Mathematics and Statistics,\\
University of Vermont, Burlington, VT 05401}
\date{\today}

\begin{abstract}

Light propagation in optical waveguides with periodically modulated
index of refraction and alternating gain and loss are investigated
for linear and nonlinear systems. Based on a multiscale perturbation
analysis, it is shown that for many non-parity-time ($\mathcal{PT}$)
symmetric waveguides, their linear spectrum is partially complex,
thus light exponentially grows or decays upon propagation, and this
growth or delay is not altered by nonlinearity. However, several
classes of non-$\mathcal{PT}$-symmetric waveguides are also
identified to possess all-real linear spectrum. In the nonlinear
regime longitudinally periodic and transversely quasi-localized
modes are found for $\mathcal{PT}$-symmetric waveguides both above
and below phase transition. These nonlinear modes are stable under
evolution and can develop from initially weak initial conditions.
\end{abstract}


\maketitle

\section{Introduction}
Parity-time ($\mathcal{PT}$)-symmetric wave systems have the
unintuitive property that their linear spectrum can be completely
real even though they contain gain and loss \cite{Bender1998}. In
spatial optics, $\mathcal{PT}$-symmetric systems can be realized by
employing symmetric index guiding and an antisymmetric gain-loss
profile \cite{Christodoulides2007,Guo2009,Segev2010}. In temporal
optics and other physical settings, $\mathcal{PT}$-symmetric systems
can be obtained as well
\cite{PT_lattice_exp,coupler1,Kottos2011,BECPT,Feng2013,Bender2013,Peng2014,Mercedeh2014}.
So far, a number of novel phenomena in optical $\mathcal{PT}$
systems have been reported, including phase transition,
nonreciprocal Bloch oscillation, unidirectional propagation,
distinct pattern of diffraction, formation of solitons and
breathers, wave blowup, and so on \cite{Bender1998,
Christodoulides2007, Guo2009, Segev2010, PT_lattice_exp, coupler1,
Feng2013, coupler2, Musslimani2008, Longhi_2009,
Musslimani_diffraction_2010, Christodoulides_uni_2011, Konotop2011,
Nixon2012, KonotopPRL, Nixon2012b, Christodoulides_uni_2012,
SegevPT}. Novel photonic devices such as $\mathcal{PT}$ lasers have
also been demonstrated \cite{Mercedeh2014}.

Research into optical $\mathcal{PT}$ systems has been largely
devoted to waveguides where the gain and loss is distributed along
the transverse direction. This leads to the natural question: what
role does $\mathcal{PT}$ symmetry play when the gain and loss is
distributed in the direction of propagation? In the study of
$\mathcal{PT}$ systems which exhibit unidirectional propagation or
Bragg solitons
\cite{Christodoulides_uni_2011,Christodoulides_uni_2012,Feng2013},
this has been touched upon. However, the models in those works
ignored the transverse effects on light propagation. For real
waveguides (i.e., without gain and loss), control of light through
modulation of the refractive index has been well documented
\cite{PeriodicPotentialReview2012}, and just recently researchers
have studied these modulations with added gain and loss distributed
in the transverse direction \cite{KivsharPsuedoPT2013}.

In this article, we study the propagation of light in complex
waveguides with periodically modulated index of refraction as well
as alternating gain and loss along the direction of propagation.
When this system is non-$\mathcal{PT}$-symmetric, we show that
linear modes often grow or decay over distance, and this growth or
decay is not affected by nonlinearity. However, several classes of
non-$\mathcal{PT}$-symmetric waveguides are found to possess
completely real linear spectrum, thus all linear modes propagate
periodically over distance. In the nonlinear regime, families of
longitudinally periodic and transversely quasi-localized solutions
exist for $\mathcal{PT}$-symmetric waveguides both below and above
phase transition. These nonlinear modes are stable under evolution
and can develop from weak initial conditions. By applying multiscale
perturbation theory, a reduced ordinary differential equation is
derived for the modes' linear and nonlinear propagation, and this
reduced model agrees well with direct simulations of the original
system.

Propagation of light in a modulated waveguide with gain and loss can
be modeled under paraxial approximation by the following nonlinear
Schr\"{o}dinger equation
\begin{equation}
\ri \psi_z  + \psi_{xx} + V(x,z) \psi + \sigma |\psi|^2 \psi = 0,
\label{e:NLS}
\end{equation}
where $z$ is the direction of propagation, $x$ is the transverse
direction, $\psi$ is the envelope function of the light's electric
field, $V(x,z)$ is a complex periodic potential whose real part is
the refractive index of the waveguide and the imaginary part
represents gain and loss, and $\sigma$ is the coefficient of the
cubic nonlinearity. A schematic diagram of our system is given in
Fig.~1. The paraxial model (\ref{e:NLS}) is valid when the waveguide
modulation is weak and the light frequency is not near the Bragg
frequency of the periodic waveguide, in which case back wave
reflection is negligible. This waveguide would be
$\mathcal{PT}$-symmetric if
\begin{equation}
V^*(x,z) = V(x, -z),
\end{equation}
where the asterisk `*' represents complex conjugation. Note that in
this $\mathcal{PT}$ condition, coordinate reflection is only in the
$z$-direction, not $x$-direction. This differs from the usual
multi-dimensional $\mathcal{PT}$ symmetry \cite{Musslimani2008} and
more resembles the partial $\mathcal{PT}$ symmetry proposed in
\cite{YangPPT}.

\begin{figure}[htbp!]
\centering
\includegraphics[width=0.45\textwidth]{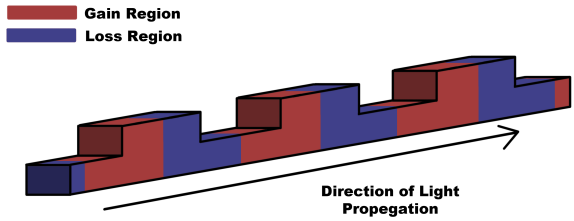}
\vspace{-0.5cm}
\caption{(Color online) A schematic diagram for an optical waveguide with modulated index of refraction and
alternating gain-loss regions. The periodic change in height represents the change in index of refraction,
while the alternating regions of red and blue represent regions of gain and loss.}
\end{figure}

To be consistent with the paraxial approximation, in this article we
consider complex waveguides where the $z$-direction modulation
appears as a small perturbation
\begin{equation}
V(x,z) = V_0(x) + \epsilon V_1(x, z),
\label{e:ExLattice1}
\end{equation}
where $V_0(x)$ is the unperturbed real refractive index which is
assumed to be localized, $\epsilon\ll 1$, and the perturbation
$V_1(z, x)$ is periodically modulated along the $z$ direction whose
period is normalized as $2\pi$. Assuming $V_1(x,z)$ has the same
transverse profile as the unperturbed index $V_0(x)$, $V_1$ then can
be expanded into a Fourier series
\begin{equation} \label{e:V1formula}
V_1(x,z) = V_0(x) \sum_{n=-\infty}^{\infty}  a_n \re^{\ri n z},
\end{equation}
where $a_n$ are complex Fourier coefficients. Without loss of
generality, we take $a_0 = 0$. The perturbed waveguide is
$\mathcal{PT}$-symmetric when all the Fourier coefficients $a_n$ are
strictly real.

\section{Multiscale perturbation analysis}

Assume that the unperturbed real waveguide $V_0$ supports a linear
discrete eigenmode $\psi = u_0(x) \re^{- \ri\mu_0z}$, where $\mu_0$
is a real propagation constant, and $u_0$ is a real localized
function satisfying
\begin{equation}  \label{e:u0}
(\partial_{xx} +V_0 +\mu_0) u_0 =0.
\end{equation}
Then in the presence of the above longitudinal waveguide
perturbations and weak nonlinearity, the perturbed solution to Eq.
(\ref{e:NLS}) can be expressed as
\begin{equation} \label{e:psi}
\psi(x,z) = u(x,z,Z) \re^{- \ri\mu_0z},
\end{equation}
where
\begin{equation}
u(x,z,Z) = \epsilon A(Z) u_0(x) + \epsilon^2 A(Z) u_1(x,z)  + \epsilon^3 U_2 + \ldots,
\label{e:EigExpansion}
\end{equation}
$A(Z)$ is a slowly varying complex envelope function, and $Z =
\epsilon^2 z$ is the slow distance variable. Substituting this
expansion into Eq.~(\ref{e:NLS}), at order $\epsilon^2$ we have
\begin{equation}
 \left(\ri \partial_z + \partial_{xx} + V_0+\mu_0 \right) u_1  = -u_0  V_1.    \nonumber
\end{equation}
Defining the operator
\begin{equation}
L_n = \partial_{xx} + V_0+\mu_0 - n,   \nonumber
\end{equation}
and expanding the solution $u_1(x,z)$ into a Fourier series
\begin{equation}
u_1(x,z) = \sum_{n= -\infty}^{\infty} u_1^{(n)}(x) \re^{\ri n z},   \nonumber
\end{equation}
each term $u_1^{(n)}(x)$ is then determined from the equation
\begin{equation}
L_n u_1^{(n)}(x) = - a_n u_0(x) V_0(x).    \nonumber
\end{equation}
Since $a_0=0$, the right hand side is zero for $n=0$. Without loss
of generality we take $u_1^{(0)}(x) = 0$ as well. Since the
potential $V_0(x)$ is localized, assuming no other discrete
eigenvalues of $V_0$ differ from $\mu_0$ by an integer, then when $n
> \mu_0$ there is no solvability condition and a localized solution
$u_1^{(n)}(x)$ is admitted. When $n < \mu_0$, however, the $L_n$
operator has non-vanishing bounded homogeneous solutions, and as a
result the corresponding solution $u_1^{(n)}$ is non-vanishing at
large $|x|$ as well if $a_n \ne 0$.

At order $\epsilon^3$ equation (\ref{e:NLS}) gives
\begin{equation}
\left(\ri \partial_z + \partial_{xx} + V_0+\mu_0 \right) U_2  = -\ri A_Z u_0 -  A u_1  V_1 -\sigma |A|^2A u_0^3.  \nonumber
\end{equation}
Decomposing the solution $U_2$ into a Fourier series in $z$,
the equation for the constant mode $U_2^{(0)}$ is found to be
\begin{equation}
L_0 U_2^{(0)} = -\ri A_Zu_0 + A \sum_{m = -\infty}^{\infty} a_{-m}a_{m} \tilde{u}_1^{(m)} V_0-\sigma |A|^2A u_0^3,  \nonumber
\end{equation}
where $L_m \tilde{u}_1^{(m)} =  u_0V_0$. In view of Eq.
(\ref{e:u0}), $u_0$ is a homogeneous solution of the above
inhomogeneous equation. Since $L_0$ is self-adjoint, in order for
this equation to be solvable, its right hand side must be orthogonal
to $u_0$. This solvability condition leads to the following ordinary
differential equation (ODE) for the evolution of the slowly-varying
envelope function $A(Z)$,
\begin{equation}
A_Z  + \ri \widetilde{\mu} A - \ri \tilde{\sigma} |A|^2 A = 0,
\label{e:Amplitude}
\end{equation}
where
\begin{equation}
\widetilde{\mu} =  \sum_{m = -\infty}^{\infty} a_{-m}a_{m} \frac{\int_{-\infty}^{\infty} V_0 u_0 \tilde{u}_1^{(m)} \rd x}
{\int_{-\infty}^{\infty} u_0^2 \rd x}, \quad
\tilde{\sigma} = \sigma \frac{\int_{-\infty}^{\infty}  u_0^4~ \rd x}{\int_{-\infty}^{\infty} u_0^2 \rd x}.
\label{e:Coefficients}
\end{equation}
This reduced ODE model will be helpful for the understanding of
linear and nonlinear dynamics of solutions in the original equation
(\ref{e:NLS}) as we will elucidate below.

First we consider the solution to the linear equation
(\ref{e:Amplitude}), i.e., with the cubic term in
(\ref{e:Amplitude}) dropped. As a concrete example we take a
waveguide where all modulations of $V_1$ are in the first harmonics,
\begin{equation} \label{e:ExPotential}
V_1(x,z) = V_0(x) \left(\re^{\ri z} + \beta \re^{-\ri z} \right),
\end{equation}
where $\beta$ is a complex constant. In this case,
$\widetilde{\mu}=\beta {c}$, where ${c}$ is a real constant
dependent on the unperturbed waveguide $V_0(x)$. Thus the linear
envelope equation (\ref{e:Amplitude}) yields
\begin{equation}
A(Z) =A_0 e^{-\ri \beta {c} Z},   \nonumber
\end{equation}
where $A_0$ is the initial envelope value. In view of Eqs.
(\ref{e:psi})-(\ref{e:EigExpansion}), this $A(Z)$ solution can be
absorbed into a shift of the eigenvalue
\begin{equation} \label{e:muformula}
\mu = \mu_0 + \epsilon^2 \beta {c}
\end{equation}
in the linear Bloch mode of Eq. (\ref{e:NLS}),
\begin{equation} \label{e:Bloch}
\psi(x,z) = \re^{-\ri \mu z} u(x,z).
\end{equation}
Then we immediately see that for non-real $\beta$ in the
first-harmonic waveguide perturbation (\ref{e:ExPotential}), a
complex eigenvalue bifurcates out from every discrete real
eigenvalue of the unperturbed waveguide. Noticing this waveguide
perturbation is $\mathcal{PT}$-symmetric when $\beta$ is real, we
conclude that the linear spectrum is partially complex when the
waveguide is non-$\mathcal{PT}$-symmetric.

Next we consider the solution to the nonlinear ODE
(\ref{e:Amplitude}). This nonlinear equation is exactly solvable,
and its general solution is
\begin{equation} \label{e:AZformula}
A(Z) = A_0 {\rm Exp} \left[ -\ri \tilde{\mu} Z -\ri \frac{ \tilde{\sigma} |A_0|^2}{2{\rm Re}[\ri\tilde{\mu}]}\left( \re^{-2{\rm Re} [\ri\tilde{\mu}] Z}-1 \right)
\right],
\end{equation}
where $A_0$ is the initial envelope value. The amplitude of this
nonlinear solution evolves as
\begin{equation} \label{e:Aabsformula}
|A(Z)|=|A_0| e^{-{\rm Re}[\ri \tilde{\mu}] Z},
\end{equation}
which is exactly the same as that of the linear solution $A(Z) =A_0
e^{-\ri \tilde{\mu} Z}$. This indicates that nonlinearity does not
affect the magnitude of the envelope solution (regardless whether it
is focusing or defocusing nonlinearity). In particular, for the
first-harmonic waveguide perturbation (\ref{e:ExPotential}) where
$\tilde{\mu}=\beta {c}$, when $\beta$ is complex with ${\rm Re}
[\ri\beta{c}]<0$, the linear solution will grow exponentially. In
this case, nonlinearity will not arrest this exponential growth at
larger amplitudes.

The above predictions for the solution dynamics are verified with
direct numerical computations of the original system (\ref{e:NLS}).
For this purpose, we take
\begin{equation}  \label{e:Vparameters1}
V_0=2\hspace{0.03cm} \mbox{sech}\hspace{0.06cm}x, \quad \epsilon=0.2
\end{equation}
in our waveguides (\ref{e:ExLattice1}) and (\ref{e:ExPotential}). In
this case, the unperturbed real waveguide $V_0$ has a single
discrete eigenvalue $\mu_0\approx -1.245$, and the parameter
${c}\approx -0.369$. Under the first-harmonic waveguide perturbation
(\ref{e:ExPotential}), we have confirmed that complex eigenvalues do
bifurcate out from $\mu_0$ in the linear spectrum according to the
formula (\ref{e:muformula}) whenever $\beta$ is non-real. In
addition, this bifurcated eigenvalue is the only complex eigenvalue
in the linear spectrum. To verify the nonlinear amplitude formula
(\ref{e:AZformula}), we choose two $\beta$ values of $\ri$ and
$-\ri$. The corresponding $z$-direction modulations of the perturbed
waveguide at $x=0$ are displayed in Fig.~2(a,b) respectively. In
these perturbed waveguides, we take the initial condition
$\psi(x,0)=\epsilon A_0 u_0(x)$, where $A_0 = 1$ and $u_0(x)$ is the
eigenmode of eigenvalue $\mu_0$ in the unperturbed waveguide $V_0$
with normalized peak height of 1. The simulation of the original
equation (\ref{e:NLS}) under this initial condition is plotted in
Fig.~2(c,d) for $\beta=\ri$ and $-\ri$ respectively. Here the
solution's amplitude at $x=0$ versus $z$ is displayed. For
comparison, the analytical amplitude solution $|\epsilon A(Z)
u_0(0)|$ with $|A(Z)|$ given by (\ref{e:Aabsformula}) is also
plotted. As predicted by the ODE model (\ref{e:Amplitude}), the
solution for $\beta=\ri$ exponentially decays, while that for
$\beta=-\ri$ exponentially grows. In the latter case, this growth is
not arrested by nonlinearity (even for longer distances than those
shown in panel (d)), in agreement with the analytical solution
(\ref{e:Aabsformula}). It is noted that amplitude oscillations in
the numerical solution are due to higher order terms in the
perturbation expansion (\ref{e:EigExpansion}), which are not
accounted for in our leading-order analytical solution plotted in
this figure. Physically these amplitude oscillations are due to
periodic gain and loss in the waveguide.

\begin{figure}[!htbp]
\centering
\includegraphics[width=0.5\textwidth]{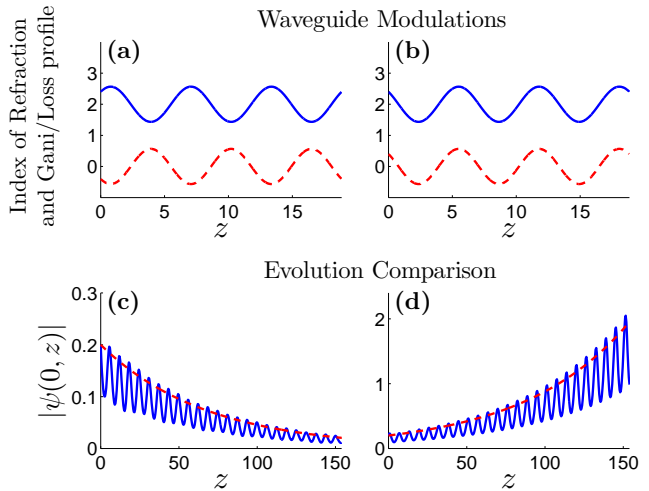}
\caption{(Color online)
(a,b) $z$-direction modulations of the waveguide (\ref{e:ExLattice1}) at $x=0$
for $\beta=\ri$ and $\beta=-\ri$ in the first-harmonic perturbations (\ref{e:ExPotential}) respectively; solid blue is
refractive-index variation and dashed red gain-loss variation (positive for gain and negative for loss);
(c,d) amplitude evolution in the nonlinear simulation of Eq. (\ref{e:NLS}) with $\sigma=1 $ for $\beta=\ri$ and $\beta=-\ri$ respectively; the
analytical solution is also plotted as dashed red lines for comparison.
}
\end{figure}

The growth and decay of solutions in Fig.~2 for different values of
$\beta$ can be intuitively understood. When Im($\beta)>0$,
modulations of the real and imaginary parts of the waveguide $V$
have a phase difference between $0$ and $\pi/2$. This means the
region with strongest index of refraction, where the pulse is at its
strongest, corresponds to a region of loss [see Fig.~2(a)], thus the
pulse decays over distance. Conversely, when Im($\beta)<0$,
modulations of the real and imaginary parts of $V$ have a phase
difference between $\pi/2$ and $\pi$. This means the region with
strongest index of refraction corresponds to a region of gain [see
Fig.~2(b)], thus the pulse grows over distance.

When first-harmonic waveguide perturbations (\ref{e:ExPotential})
are $\mathcal{PT}$-symmetric, i.e., $\beta$ is real, the perturbed
waveguide can be rewritten as
\begin{equation} \label{e:VPT}
V(x,z)=V_0(x)\left[1+\epsilon(1+\beta)\cos z+\ri\epsilon(1-\beta)\sin z\right],
\end{equation}
thus $1-\beta$ controls the strength of gain and loss. A common
phenomenon in $\mathcal{PT}$-symmetric systems is that when the
gain-loss strength (relative to the real refractive index) exceeds a
certain threshold, complex eigenvalues can appear in the linear
spectrum --- a phenomenon called phase transition
\cite{Bender1998,Guo2009,
Segev2010,Musslimani2008,PT_lattice_exp,coupler1,Kottos2011,BECPT,Feng2013,Bender2013,Peng2014,Mercedeh2014,Nixon2012,
KonotopPRL}. Then for $\mathcal{PT}$-symmetric waveguides
(\ref{e:VPT}), can phase transition occur when $\beta$ is varied?
For the previously chosen function $V_0(x)$ and $\epsilon$ value in
(\ref{e:Vparameters1}), our numerics did not detect phase
transition. But for some other choices of $V_0(x)$ and $\epsilon$,
phase transition can indeed occur. For instance, when we choose
\begin{equation}  \label{e:Vparameters2}
V_0=2\mbox{sech}^2x, \quad \epsilon=0.2,
\end{equation}
phase transition occurs at $\beta=0$. The linear spectrum is
all-real when $\beta>0$ and becomes complex when $\beta<0$. Two
waveguides with $\beta$ values of $0.5$ and $-0.5$ (below and above
phase transition) are illustrated in Fig.~3(a,b) respectively. It
should be pointed out that even though this $\mathcal{PT}$-symmetric
waveguide is above phase transition when $\beta<0$, the complex
eigenvalues in its linear spectrum have very small imaginary parts,
meaning that the growth or decay of the corresponding linear
eigenmodes is very weak. For instance, when $\beta=-0.5$, the
complex eigenvalue with maximal imaginary part is $\mu =-0.9864 +
0.0026\ri$. Notice that the real part of this complex eigenvalue is
very well predicted by the analytical formula (\ref{e:muformula}),
which gives $-0.9862$ since $\mu_0=-1$ and ${c}\approx -0.691$ for
the present waveguide. But the imaginary part of this complex
eigenvalue is not captured by the perturbation analysis of this
section since it is very small and occurs at higher orders of
perturbation expansions.

\section{Non-$\mathcal{PT}$-symmetric waveguides with all-real
linear spectrum}

It is seen from the previous section that for first-harmonic
waveguide perturbations (\ref{e:ExPotential}), all-real linear
spectra are possible only for real values of $\beta$, i.e., when the
perturbed waveguide is $\mathcal{PT}$-symmetric. However, we have
found two notable families of complex waveguides which are
non-$\mathcal{PT}$-symmetric but still possess all-real linear
spectra. This is quite surprising, since in complex waveguides with
transverse gain-loss variations, all-real spectra are very rare for
non-$\mathcal{PT}$-symmetric systems \cite{SuperSymmetry2013}.

The first family consists of waveguides
(\ref{e:ExLattice1})-(\ref{e:V1formula}) with unidirectional Fourier
series decomposition, i.e., $a_n = 0$ for either $n<0$ or $n>0$. In
our calculation of the shifted eigenvalue $\mu = \mu_0 + \epsilon^2
\tilde{\mu}$ with $\tilde{\mu}$ given in Eq. (\ref{e:Coefficients}),
notice that $\tilde{\mu}=0$ for a unidirectional Fourier series,
hence the eigenvalue $\mu_0$ does not shift at all under these
complex waveguide perturbations. Regarding other eigenvalues in the
linear spectrum, we have verified numerically that they do not shift
to the complex plane either, thus the linear spectrum is all-real
for waveguides of this type.

The second family consists of separable waveguides, $V(x,z) = V_a(z)
+ V_b(x)$, where $\int_0^{2\pi} \Imag[V_a(z)]\rd z=0$ (meaning that
the gain and loss are balanced along the propagation direction), and
$V_b(x)$ is real. In this case, Bloch modes (\ref{e:Bloch}) in the
linear equation (\ref{e:NLS}) can be decomposed as $\mu =
\mu_a+\mu_b$ and $u(x,z) = u_a(z)u_b(x)$, where $(u_a, \mu_a), (u_b,
\mu_b)$ satisfy the following one-dimensional eigenvalue problems
\begin{align*}
& [\ri \partial_z  + V_a(z)]u_a = -\mu_a u_a,   \\
& [\partial_{xx} + V_b(x)] u_b =-\mu_b u_b.
\end{align*}
The first eigenvalue problem has an exact solution
$$u_a(z) = {\rm Exp}\left\{ \ri \mu_a z  + \ri\int_0^{z} V_a(\xi)\rd \xi \right\}.$$
Thus for waveguides with equal amounts of gain and loss, i.e.,
$\int_0^{2\pi} \Imag[V_a(z)]\rd z=0$, its $\mu_a$-spectrum is
all-real. The second eigenvalue problem is a Schr\"{o}dinger
eigenvalue problem. Thus for real waveguides $V_b(x)$, its spectrum
is also all-real. Together, we see that for the separable waveguides
of the above form, the linear spectrum is all-real. Notice that
these separable waveguides are non-$\mathcal{PT}$-symmetric in
general, thus they constitute another large class of
non-$\mathcal{PT}$-symmetric waveguides with all-real spectra.

\section{Longitudinally-periodic nonlinear modes}

In this section we consider nonlinear $z$-periodic modes in these
modulated waveguides. Such modes are of the form
\begin{equation} \label{e:psisoliton}
\psi(x,z)=e^{-\ri\mu z} u(x,z),
\end{equation}
where $\mu$ is a real propagation constant, and $u(x,z)$ is
$2\pi$-periodic in $z$. From the reduced ODE model
(\ref{e:Amplitude}) for the first-harmonic waveguide perturbation
(\ref{e:ExPotential}), we see that when the waveguide is
non-$\mathcal{PT}$-symmetric (i.e., $\beta$ is non-real) the
solution (\ref{e:AZformula}) will either grow or decay (since
$\tilde{\mu}=\beta {c}$ is complex), thus nonlinear $z$-periodic
modes are not expected. But when the waveguide is
$\mathcal{PT}$-symmetric (with real $\beta$), the nonlinear solution
to the ODE model is
\begin{equation} \label{e:AZformulaPT}
A(Z) = A_0 {\rm Exp} \left[ -\ri \beta{c} Z +\ri \tilde{\sigma} |A_0|^2 Z\right].
\end{equation}
Since $\beta{c}$ and $\tilde{\sigma}$ are real, when this
$A(Z)$ function is substituted into the perturbation series solution
(\ref{e:psi})-(\ref{e:EigExpansion}), analytical $z$-periodic
nonlinear modes (\ref{e:psisoliton}) with
\begin{equation}  \label{e:mubifurcation}
\mu=\mu_0+\epsilon^2 (\beta {c} -\tilde{\sigma}|A_0|^2)
\end{equation}
are then obtained. In this $\mu$ formula, the amplitude parameter
$A_0$ is arbitrary. Thus a continuous family of nonlinear
$z$-periodic modes parameterized by the propagation constant $\mu$
are predicted. Our perturbation analysis also reveals another
important property about these $z$-periodic modes, i.e., they
contain weak transversely nonlocal tails and are thus not fully
localized. The order at which these nonlocal tails appear in the
perturbation series depends on the unperturbed waveguide $V_0(x)$ as
well as the waveguide perturbation $V_1(x)$. For the first-harmonic
perturbation (\ref{e:ExPotential}) and $V_0=2\mbox{sech}x$ [as in
(\ref{e:Vparameters1})], these nonlocal tails appear at the
$O(\epsilon^3)$ term in the $e^{-2\ri z}$ harmonics. For
perturbations with $V_0=2\mbox{sech}^2x$ [as in
(\ref{e:Vparameters2})],  these nonlocal tails appear at the
$O(\epsilon^2)$ term in the $e^{-\ri z}$ harmonics. Since these
transversely nonlocal tails occur at higher orders of the
perturbation series, the resulting $z$-periodic nonlinear mode is
then quasi-localized, i.e., the height of the solution's tails at
$x\rightarrow \pm \infty$ is much less than the solution's peak
amplitude.

\begin{figure}[!htbp]
    \centering
    \includegraphics[width=0.45\textwidth]{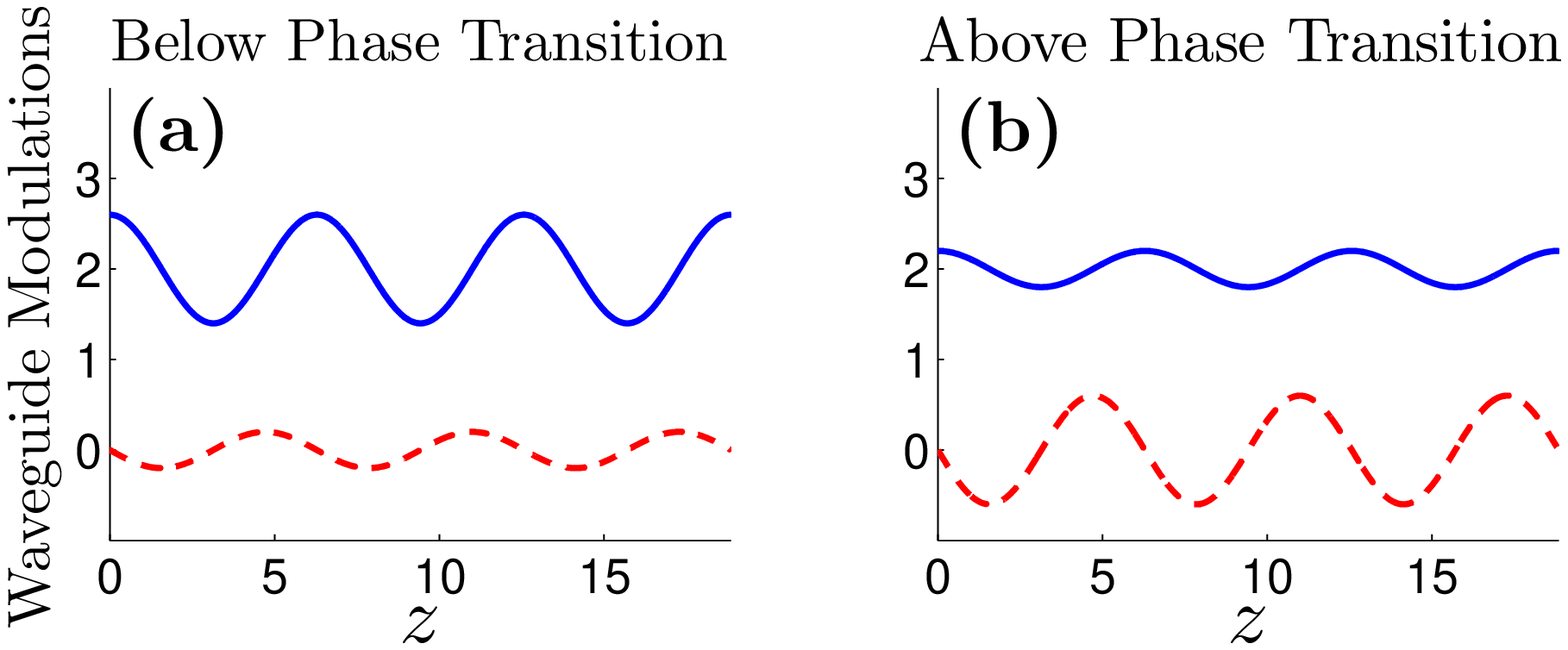}

    \vspace{0.4cm}
    \includegraphics[width=0.45\textwidth]{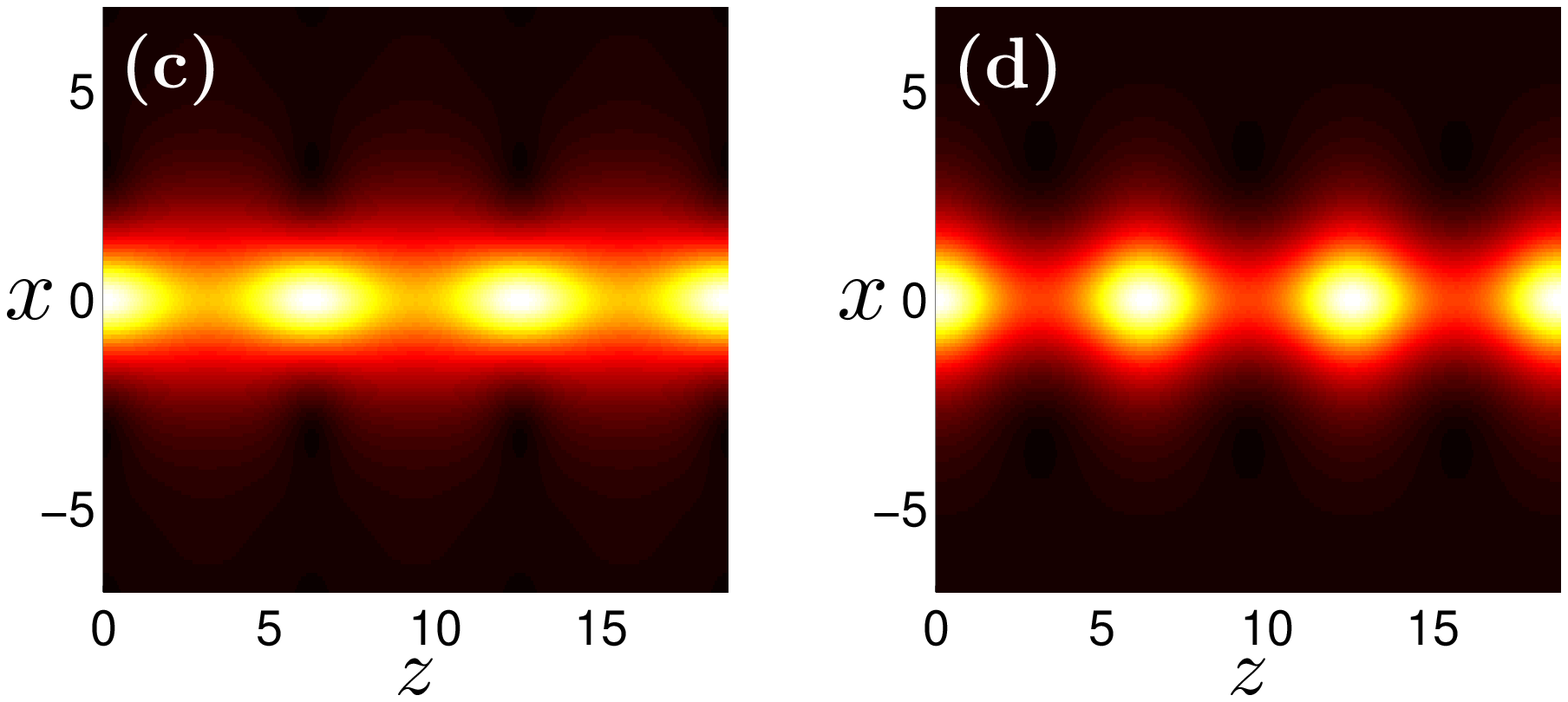}

     \vspace{0.cm}
     \includegraphics[width=0.45\textwidth]{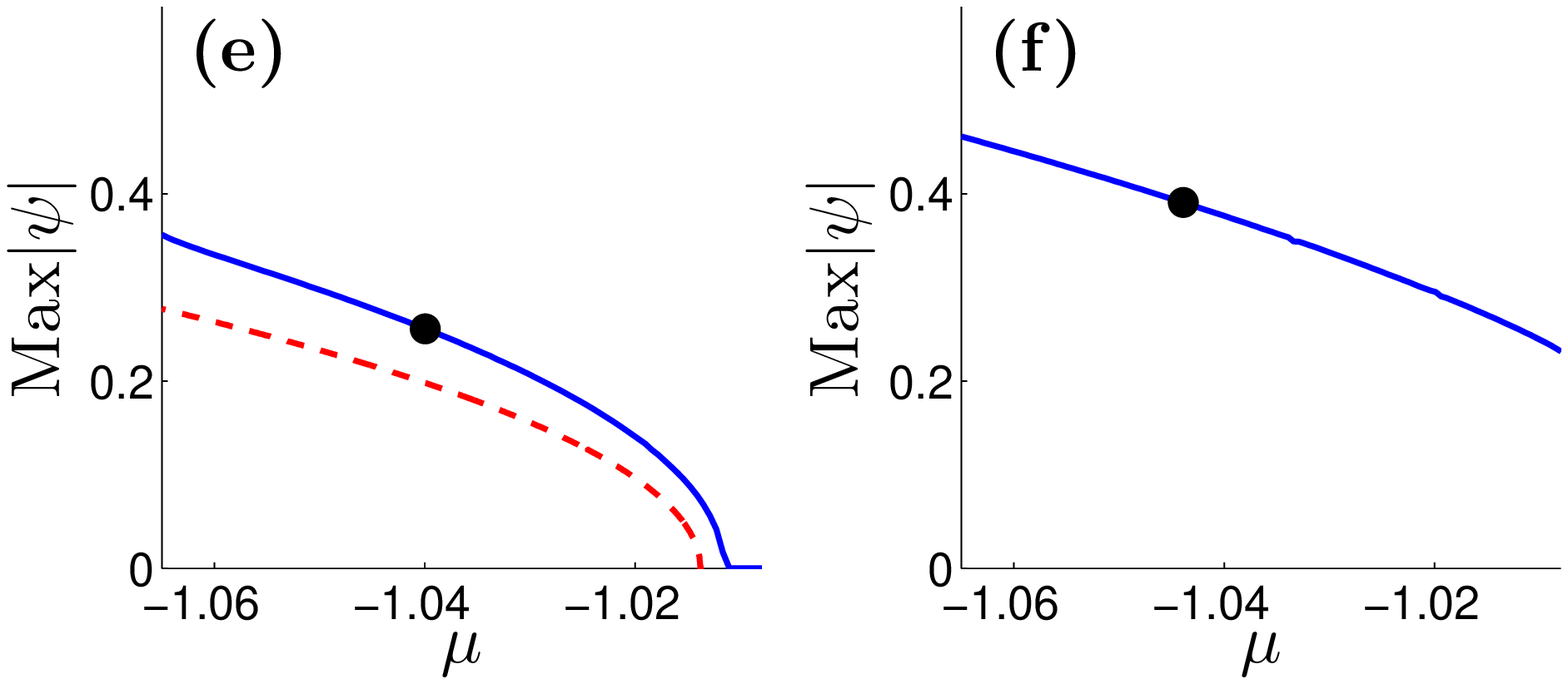}
    \caption{(Color online) Families of longitudinally periodic and transversely quasi-localized nonlinear modes in $\mathcal{PT}$-symmetric
    waveguides below and above phase transition. The waveguide is (\ref{e:ExLattice1}) with first-harmonic perturbations (\ref{e:ExPotential}),
    where $V_0=2\mbox{sech}^2x$, $\epsilon=0.2$ and $\sigma=1$.
    (a,b) Modulated waveguides versus $z$ at $x=0$ for $\beta =0.5$ (below phase transition) and $\beta=-0.5$ (above phase
    transition) respectively;  solid blue is refractive-index variation and dashed red gain-loss variation (positive for gain and negative for loss);
    (c,d) example nonlinear modes in waveguides of (a,b)
    respectively; (e,f) nonlinear modes' peak amplitude versus the propagation constant
     $\mu$ in waveguides of (a,b); solid blue lines are numerical values, while the
     dashed red line in (e) is analytical predictions. The locations of example modes in (c,d) are marked by black dots.   }
\end{figure}

Numerically we have confirmed the existence of these $z$-periodic
and transversely quasi-localized nonlinear modes in Eq.
(\ref{e:NLS}) for $\mathcal{PT}$-symmetric waveguides. In addition,
we have found that these modes exist both below and above phase
transition. These solutions are computed as a boundary value problem
in the $(x,z)$ space by the Newton-conjugate-gradient method
\cite{JiankeNCG}. To demonstrate, we take the first-harmonic
perturbation (\ref{e:ExPotential}) with $V_0$ and $\epsilon$ given
in Eq. (\ref{e:Vparameters2}). We also take $\sigma=1$ (focusing
nonlinearity). For $\beta=0.5$ below phase transition, this solution
family is displayed in Fig.~3(e). It is seen that these modes
bifurcate from $\mu\approx -1.011$ where its amplitude approaches
zero. The analytical bifurcation point from formula
(\ref{e:mubifurcation}), with $A_0$ set to zero, is
$\mu_{anal}\approx -1.014$, since $\mu_0=-1$ and ${c}\approx -0.691$
for the present waveguide. Apparently the numerical and analytical
bifurcation points are in good agreement. Further comparison between
the numerically obtained peak amplitudes of these modes and
analytically obtained $\epsilon A_0$ values from Eq.
(\ref{e:mubifurcation}) for varying $\mu$ values can be seen in
Fig.~3(e). An example solution, with $\mu=-1.04$, is shown in
Fig.~3(c). Notice that this solution is strongly localized, since
its nonlocal transverse tails are very weak and thus almost
invisible.

At $\beta=-0.5$ above phase transition, these nonlinear modes are
found as well, whose peak amplitude versus the propagation constant
$\mu$ is depicted in Fig.~3(f). These solutions do not bifurcate
from infinitesimal linear modes, thus its peak amplitude does not
reach zero. An example solution at $\mu=-1.044$ is shown in
Fig.~3(d). This solution is also strongly localized with tails
almost invisible.

\begin{figure}[!htbp]
    \centering
    \includegraphics[width=0.45\textwidth]{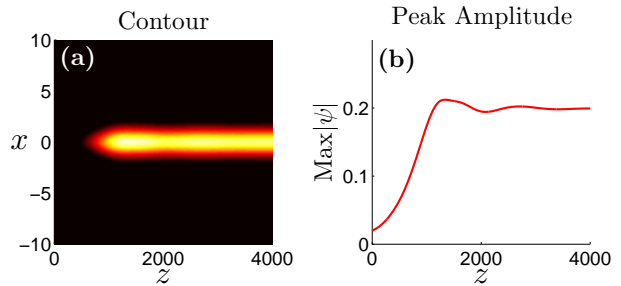}
    \caption{(Color online) Evolution of a weak initial condition in a $\mathcal{PT}$-symmetric waveguide above phase transition as shown in Fig.~3(b).
    (a) Solution evolution in the $(x,z)$ plane; (b) amplitude evolution versus distance $z$. In this figure, solutions are plotted at distances
    $z=2n\pi$, where $n$ is an integer, thus amplitude variations inside each $z$-period $2\pi$ (caused by gain and loss regions) is invisible.}
\end{figure}

We have examined the stability of these $z$-periodic nonlinear modes
by simulating their evolution under perturbations in
Eq.~\eqref{e:NLS}, and they are found to be stable. This stability
holds even when the waveguide is above phase transition. In the
latter case, an initial localized function whose amplitude is above
the threshold of periodic nonlinear modes in Fig.~3(f) would evolve
into one of these modes. If the initial amplitude is very small,
then it will first grow exponentially due to the existence of
complex (unstable) eigenvalues in the linear spectrum above phase
transition. Subsequently in the nonlinear regime, we find that its
growth saturates, and the solution approaches a $z$-periodic
nonlinear state. This evolution is illustrated in Fig.~4 for the
waveguide of Fig.~3(b) (with $\sigma = 1$) under the initial
condition $\psi(x,0)=0.02\hspace{0.04cm} \sech x$. This growth
saturation by nonlinearity in $\mathcal{PT}$-symmetric waveguides
(above phase transition) contrasts that in
non-$\mathcal{PT}$-symmetric waveguides, where the linear
exponential growth is not arrested by nonlinearity [see Sec. II and
Fig.~2(d)]. This growth saturation occurs over a long distance
though, since the growth rates of linear modes are very small (see
the end of Sec. II).

\section{Summary}

In summary, we have studied light propagation in complex waveguides
with periodic refractive index modulations and alternating gain and
loss along the direction of propagation. Our analysis is based on a
multi-scale perturbation theory, supplemented by direct numerical
simulations. We have shown that non-$\mathcal{PT}$-symmetric
waveguides often possess complex eigenvalues in their linear
spectrum, but several classes of such waveguides with all-real
linear spectra are also identified. In the nonlinear regime, we have
shown that for non-$\mathcal{PT}$-symmetric waveguides, cubic
nonlinearity does not alter the exponential growth or decay of the
related linear system. But for $\mathcal{PT}$-symmetric waveguides,
continuous families of longitudinally periodic and transversely
quasi-localized nonlinear modes exist both below and above phase
transition. In the latter case, low-amplitude initial conditions
eventually develop into these nonlinear periodic states.

\section*{Acknowledgment} This work was supported in part by the Air Force Office of
Scientific Research (USAF 9550-12-1-0244) and the National Science
Foundation (DMS-1311730).

\end{document}